\documentclass[a4paper,11pt]{article}
\pdfoutput=1
\usepackage{jheppub}
\usepackage{amssymb,amsmath,mathrsfs,enumerate}
\usepackage{graphicx,rotate,multicol}
\usepackage{float}

\usepackage{subfig}

\usepackage{slashed}
\usepackage{mathtools}
\usepackage{multirow}

\allowdisplaybreaks

\title{\boldmath Primordial Black Holes and Gravitational Waves in the $U(1)_{B-L}$ Extended Inert Doublet Model: A First-Order Phase Transition Perspective}

\author[a]{Indra Kumar Banerjee,}
\author[a]{Ujjal Kumar Dey,}
\author[b]{and Shaaban Khalil}
\affiliation[a]{Department of Physical Sciences, Indian Institute of Science Education and Research Berhampur,\\Transit Campus, Government ITI, Berhampur 760010, Odisha, India}
\affiliation[b]{Center for Fundamental Physics, \\
Zewail City of Science and Technology, 6th of October City, Giza
12578, Egypt}

\emailAdd{indrab@iiserbpr.ac.in}
\emailAdd{ujjal@iiserbpr.ac.in}
\emailAdd{skhalil@zewailcity.edu.eg}

\abstract{
We conduct an analysis of a $U(1)_{B-L}$ extended inert doublet model and obtained the parameter space allowing strong first order phase transitions. We show that a large part of the parameter space can cause double first-order phase transitions. Whereas both of these phase transitions can generate a detectable stochastic gravitational wave background, one of them can create primordial black holes with appreciable abundance. The primordial black holes generated at the high scale transition can account for the dark matter maintaining the correct relic abundance. We also show specific benchmark cases and their consequences from the aspect of primordial black holes and gravitational waves.}

\begin{document}
\maketitle
\flushbottom

\section{Introduction}
\label{sec:intro}

Primordial Black Holes (PBHs) and Gravitational Waves (GWs) stand as cosmic messengers, shedding light on the foundational truths of the universe. Their presence and properties serve as windows into the early universe, revealing tantalizing clues about its evolution and composition.  PBHs can be formed  from the collapse of high-density inhomogeneities in the early universe through various mechanisms, including a strong first-order phase transition, characterized by the nucleation and collision of true-vacuum bubbles. This type of strong first-order phase transition also contributes to the generation of GWs.

The idea of PBHs was initially proposed in the 1960s by Zel'dovich and Novikov~\cite{Zeldovich:1967lct} and later refined by Hawking and Carr in the 1970s \cite{Carr:1974nx}. They suggested that PBHs could have emerged from dense regions in the early universe in a different way than the ones we typically see today, which are formed from collapsing massive stars. Direct observational evidence of PBHs' existence remains challenging. Various indirect methods, such as gravitational lensing effects~\cite{Niikura:2017zjd, Griest:2013esa, Niikura:2019kqi, Macho:2000nvd, EROS-2:2006ryy, Zumalacarregui:2017qqd} or their potential influence on cosmic microwave background radiation~\cite{Poulin:2017bwe} etc. have been proposed to detect PBHs. Ongoing research continues to explore these mysterious objects and their potential implications for our understanding of the universe. The detection of gravitational waves from black hole mergers by LIGO and Virgo sparked renewed interest in PBHs~\cite{LIGOScientific:2016aoc}. Furthermore, recent developments in Pulsar Timing Arrays (PTAs) have presented strong evidence for a stochastic gravitational wave background at nano-Hertz frequencies. This background is thought to be originated from primordial sources \cite{NANOGrav:2023gor, NANOGrav:2023hde, EPTA:2023fyk, EPTA:2023sfo, EPTA:2023xxk, Reardon:2023gzh, Zic:2023gta, Reardon:2023zen, Xu:2023wog, Ellis:2023oxs}.
Moreover, PBHs are considered as a potential explanation for Dark Matter (DM). PBHs can account for certain observations, such as gravitational lensing, without introducing new particles. However, this depends on their mass falling within a specific range and their relic density being constrained by observational limits.

Numerous formation mechanisms for PBHs have been explored, including collapse from density fluctuations during different cosmic eras~\cite{Carr:1974nx, Grillo:1980rt, Khlopov:1980mg, Khlopov:1985fch, Carr:1994ar, Chakraborty:2022mwu}, collapse in single and multi-field inflationary models~\cite{Carr:1993aq,  Bullock:1996at, Saito:2008em, Kawai:2021edk, Kawai:2021bye, Choudhury:2023hvf, Randall:1995dj, Garcia-Bellido:2016dkw, Braglia:2020eai, Kawai:2022emp, Heydari:2021gea, Heydari:2021qsr}, collapse of cosmic string loops~\cite{Hawking:1987bn, Borah:2023iqo}, and collapse during a First Order Phase Transition (FOPT)~\cite{Crawford:1982yz, Kawana:2021tde, Baker:2021nyl, Huang:2022him, Kawana:2022olo, Liu:2021svg, Gouttenoire:2023naa, Lewicki:2023ioy, Roshan:2024qnv, Kanemura:2024pae, Goncalves:2024vkj,Banerjee:2023qya,Lewicki:2024ghw}. Many of these processes have also been considered as potential sources of GWs~\cite{Matarrese:1993zf, Matarrese:1996pp, Matarrese:1997ay, Khlebnikov:1997di, Easther:2006vd, Easther:2007vj, Choudhury:2013woa, Binetruy:1981mr, Vachaspati:1984gt, Hindmarsh:1994re, Witten:1984rs, Hogan:1986dsh}. We will focus on the FOPT origin of PBHs and GWs within the framework of the $U(1)_{B-L}$ extension of the Inert Doublet Model (IDM), where the FOPT is associated with the high-scale breaking of the $U(1)_{B-L}$ symmetry as well as electroweak symmetry breaking.

As a simple inert doublet extension of the Standard Model (SM) IDM was proposed to address the dark matter issue under the WIMP paradigm. It was soon realised that with the introduction of right handed neutrinos this model can explain the active neutrino masses radiatively~\cite{Deshpande:1977rw, Ma:2006fn, Ma:2006km, Barbieri:2006dq, LopezHonorez:2006gr}.
However, the discovery of Higgs boson at 125 GeV and the ever-increasing precision of direct detection experiments excluding so much of the parameter space takes away the charm of even the neutrino augmented IDM. 
To address this we consider an additional $U(1)_{B-L}$ symmetry which can allow a significant first-order phase transition (FOPT) that can generate PBHs and GWs~\cite{Baldes:2023rqv, Salvio:2023ynn}. As a consequence these PBHs can play the role of DM and the burden of satisfying the DM constraints by the inert scalar is lifted. Additionally this allows us a complementary probe of the scenario by means of GW astronomy.
The paper is organized as follows. In section~\ref{sec:model} we briefly overview the $U(1)_{B-L}$ extension of the IDM. In section~\ref{sec:fopt} we study the FOPT with $U(1)_{B-L}$ as well as the electroweak symmetry breaking. We then discuss briefly the creation of gravitational waves and the PBHs out of the FOPT. Afterwards in section~\ref{sec:res} we discuss our results in terms of the parameter space relevant for experimental observations. Finally we summarise and conclude in section~\ref{sec:concl}. 

\section{$U(1)_{B-L}$ Extension of IDM}
\label{sec:model}
In this section, we delve into the particle content and the  spontaneous symmetry breaking within the context of the $U(1)_{B-L}$ extension of the Inert Doublet Model (IDM). The vanilla IDM is an extension of the SM of particle physics that introduces an additional scalar doublet beyond the one already present in the SM. The IDM is motivated by several theoretical and phenomenological considerations, including the desire to explain DM, the stability of the Higgs potential etc.  The $U(1)_{B-L}$ extension of the IDM introduces an additional gauge symmetry, where $B$ stands for baryon number and $L$ stands for lepton number. The anomaly cancellation condition associated with the $U(1)_{B-L}$ gauge symmetry necessitates the presence of three right-handed neutrinos, denoted as $N_R$. This extension provides a framework for understanding the origin of neutrino masses through the seesaw mechanism. The Higgs sector of this model should contain one SM singlet complex scalar field $\chi$ that can spontaneously break the $U(1)_{B-L}$ symmetry. Additionally, there are two $SU(2)_L$ doublets: $\Phi_1$, with even $\mathbb{Z}_2$ discrete symmetry, that breaks the $SU(2)_L \times U(1)_Y$ symmetry down to $U(1)_{\mathrm{em}}$, and $\Phi_2$, with odd $\mathbb{Z}_2$ discrete symmetry, that acts as inert doublet. The quantum numbers of the BSM fields corresponding to the different symmetry groups are shown in Tab.~\ref{tab:part}. 
\begin{table}[H]
\centering
\begin{tabular}{|c|c|c|c|}
\hline
BSM fields & $SU(2)_{L}$ & $U(1)_Y$ & $U(1)_{B-L}$ \\ \hline \hline
$\chi$     & 1           & 0        & 2            \\ \hline
$\Phi_2$   & 2           & 1        & 0            \\ \hline
$N_{R_i}~(i=1,2,3)$   & 1           & 0        & $-1$           \\ \hline
\end{tabular}
\caption{The quantum numbers of the BSM fields corresponding to the symmetry groups.}
\label{tab:part}
\end{table}
Before going to the scalar part of the Lagrangian, we would like to discuss the relevant Yukawa part which is as follows,
\begin{align}
\mathcal{L}_{\mathcal{Y}}=-\left(Y_{N^1_{ij}}\overline{L_{i}}\tilde{\Phi}_1 N_{R_j}+Y_{N^2_{ij}}\overline{L_{i}}\tilde{\Phi}_2 N_{R_j}-\dfrac{1}{2}y_i\overline{N^C_{R_i}}\chi N_{R_i}+\mathrm{h.c.}\right),
\label{eq:yukawa}
\end{align}
where $\tilde{\Phi}_{1,2}=i\sigma_2\Phi_{1,2}^*$, $\sigma_2$ is the second Pauli matrix, and $L_i = (\nu_i,l_i)$ is the SM lepton doublet. The breaking of the $U(1)_{B-L}$ symmetry generates the mass matrix $M_{R_i}$ of the right handed neutrinos which takes the form $M_{R_i}=y_i v_{\chi}$, where $v_{\chi}$ is the vacuum expectation value (VEV) acquired by $\chi$. There can be ways through which neutrino mass can be generated. For example, only the first and the third term of the RHS of Eq.~\eqref{eq:yukawa} can lead to the mass generation for the active neutrinos through a type-I seesaw mechanism. In that case, $\mathcal{O}(Y_{N^1})$ has to be $\mathcal{O}(10^{-4})$ for $m_{\nu}\sim 0.1\mathrm{~eV}$ for the parameter space we consider. On the other hand, only the second and third term in the RHS of can lead to active neutrino mass generation through radiative seesaw mechanism. In that case, $\mathcal{O}(Y_{N^2})$ has to be $\mathcal{O}(10^{-8})$ for $m_{\nu}\sim 0.1\mathrm{~eV}$ for the parameter space we consider. It is worth mentioning here that although we provide an estimation, we do not strive to fit the neutrino oscillation data since that is outside the scope of this work.

In the scalar sector we assume that the model has classical conformal invariance. Under this invariance -- in the absence of dimensionful parameters -- the scalar potential for two Higgs doublets ($\Phi_{1,2}$) and a singlet scalar (\( \chi \)) can be written as,
\begin{align}
V(\Phi_1, \Phi_2, \chi) = &\lambda_1 \vert \Phi_1\vert^4 + \lambda_2 \vert \Phi_2 \vert^4 +  \lambda_3 \vert\Phi_1\vert^2 \vert\Phi_2\vert^2 +  \lambda_4 \vert \chi \vert^4 +  \lambda_5 \vert \Phi_1^\dagger \Phi_2 \vert^2  \nonumber\\
&+ \lambda_6 \left[(\Phi_1^\dagger \Phi_2)^2 + \text{h.c.}\right]  +\  \lambda_7 \vert \chi \vert^2  \vert \Phi_1\vert^2 + \lambda_8  \vert \chi \vert^2  \vert \Phi_2\vert^2 ,
\label{Vtree}
\end{align}
where $( \lambda_1,\ldots, \lambda_8 )$ are dimensionless coupling constants which determine the strength of the interactions.  The scalar field $\chi$ is unable to acquire a non-vanishing VEV solely from the tree-level potential, as represented by the quartic term $\lambda_4|\chi|^4$. This limitation implies that the tree-level potential alone can not induce spontaneous $U(1)_{B-L}$ symmetry breaking. However, additional contributions, such as those from the one-loop Coleman-Weinberg potential, can facilitate the generation of a non-zero VEV for $\chi$. The Coleman-Weinberg potential at one-loop order arises from quantum corrections to the scalar potential due to loop diagrams involving the scalar field $\chi$ and other fields in the theory. The general expression for the one-loop Coleman-Weinberg potential can be written as,
\begin{align}
V_{\text{CW}}(\phi) = (-1)^{2s_i}g_i\dfrac{m_i^4(\phi)}{64\pi^2}\left[\log\left(\dfrac{m_i^2(\phi)}{\Lambda^2}\right)-c_i\right],
\label{CW}
\end{align}
where $m_i(\phi)$ is the field-dependent masses of the different particle species in the model, $\phi$ is the classical background field associated with the field $\chi$, and $\Lambda$ is the renormalization scale which we take to be of the order of the VEV realised by the $\chi$ field. In the previous equation $s_i$ and $g_i$ are the spin and the degree of freedom of the $i$th particle, respectively, whereas $c_i$ are constants which in the $\overline{\text{MS}}$ scheme takes the value $3/2$ for scalars and fermions and $5/6$ for gauge bosons. The Coleman-Weinberg potential typically includes contributions from Yukawa, gauge, and quartic couplings of the fields involved in the model. This expression captures the quantum corrections to the scalar potential due to loop diagrams and provides insights into the possibility of spontaneous symmetry breaking and the generation of non-zero VEVs for scalar fields like $\chi$.
It is also worth mentioning here that the nature of the transition which breaks different symmetries, i.e. whether they  are of first order, second order or a cross-over in nature, depends of the strength of the interactions. Furthermore, in order for the right-handed neutrinos to be sufficiently heavy, we consider the VEV of the field $\chi$ to be much larger than the VEV of the SM Higgs doublet $\Phi$, which makes the analysis of the breaking of the $U(1)_{B-L}$ effectively driven by a single field, i.e. $\chi$.


\section{First Order Phase Transition}
\label{sec:fopt}
In order to study the details of the singlet scalar driven FOPT associated with $U(1)_{B-L}$ symmetry breaking in conformal $U(1)_{B-L}$ extension IDM model, we first express the full one-loop corrected potential $V_{\text{tot}}$ as follows,
\begin{align}
V_{\text{tot}} = V_{\text{tree}} + V_{\text{CW}} + V_{\text{th}} ,
\end{align}
where $V_{\text{tree}}$ represents the tree-level potential, as defined in Eq.~(\ref{Vtree}) whereas $ V_{\text{CW}} $ is defined in Eq.~(\ref{CW}). 
Finally, the thermal potential $V_{\text{th}}$ is given by,
\begin{align} 
V_{\text{th}}(\phi,T) = \frac{T^4}{2\pi^2}
	\left[ 
	\sum_{i=\text{bosons}} 
	J_{B}\left(\frac{m_i(\phi)}{T}\right) 
	+ \sum_{i=\text{fermions}} 
	J_{F}\left(\frac{m_i(\phi)}{T}\right)
	\right] 
	+ V_{\mathrm{D}}(\phi,T),
\end{align}
where the sums run over all bosonic and fermionic degrees of freedom, and $J_B $ and $ J_F $ are thermal loop functions that depend on the mass of the particles and the temperature $ T $. Here, $V_{\mathrm{D}}$ is the contribution from the daisy resummation and can be expressed as,
\begin{align}
V_{\mathrm{D}}(\phi,T)=-\sum_{i}\dfrac{g_iT}{12\pi}\left[m_i^3(\phi,T)-m_i^3(\phi)\right],
\end{align}
where $m_i^2(\phi,T)=m_i(\phi)+\Pi_i$ with $\Pi_i$ being the thermal correction to the mass of the $i$-th boson species.
As the temperature of the universe decreases, this temperature-dependent potential acquires a non-zero global minima (true vacuum). The temperature, at which the false vacuum (pre-existing local minima) and the true vacuum becomes degenerate, is called the critical temperature ($T_c$). As the temperature lowers further down, the universe shifts to the true vacuum through the nucleation of true vacuum bubble. The nucleation rate of the true vacuum bubbles can be expressed as,
\begin{align}
\Gamma(T)=\mathcal{A}(T)e^{-S_{3}(T)/T},
\end{align}
where $\mathcal{A}(T)$ is a numerical prefactor which is $\sim T^4$ and $S_{3}(T)$ is the three-dimensional Euclidean action of the configuration. This action can be obtained from the expression,
\begin{align}
S_3=\int^{\infty}_{0}dr 4\pi r^2\left[\dfrac{1}{2}\left(\dfrac{d\phi}{dr}\right)^2+V_{\mathrm{tot}}(\phi,T)\right].
\end{align}
The temperature at which the true vacuum bubbles nucleate is called the nucleation temperature $(T_n)$ which can be defined from the idea that at this temperature, the nucleation rate per unit comoving volume is $\mathcal{O}(1)$, i.e.
\begin{align}
\Gamma(T_n)=H^4(T_n),
\end{align}
where $H(T)$ is the Hubble parameter of the universe at temperature $T$ and can be approximated as $H(T)\approx 1.66\sqrt{g_*}T^2/M_{\mathrm{Pl}}$ where $g_*$ is the effective degrees of freedom of the universe at temperature $T$ and $M_{\mathrm{Pl}}$ is the Plank mass. Furthermore, the temperature at which $34\%$ of the comoving volume is in true vacuum is called percolation temperature, $T_p$. This is equivalent to the temperature at which the probability of finding a point in the comoving volume which is still in the false vacuum is 0.71. This probability can be expressed as,
\begin{align}
\mathcal{P}(T)=e^{-\mathcal{I}(T)},
\end{align}
where $\mathcal{I}(T)$ is defined as,
\begin{align}
\mathcal{I}(T)=\dfrac{4\pi}{3}\int^{T_c}_T \dfrac{dT'}{{T'}^4}\dfrac{\Gamma(T')}{H(T')}\left(\int^{T'}_T \dfrac{dT''}{H(T'')}\right)^3.
\end{align}
It is worth mentioning that the strength of a FOPT can be qualitatively understood in a few different ways. If $v_c/T_c>1$ and $v_n/T_n>1$, then the FOPT is said to be a strong FOPT, where $v_c$ and $v_n$ are the finite temperature minima of the field driving the FOPT at $T_c$ and $T_n$ respectively. Furthermore, the ratio between the energy released in the form of latent heat with the radiation energy density at the percolation temperature is also a measure of the strength of the FOPT which is expressed as,
\begin{align}
\alpha=\dfrac{1}{\rho_{\mathrm{rad}}(T_p)}\left[\Delta V_{\mathrm{tot}}-\dfrac{T}{4}\dfrac{d}{dT}(\Delta V_{\mathrm{tot}})\right]_{T=T_p},
\end{align}
where $\Delta V_{\mathrm{tot}}=V_{\mathrm{tot}}\vert_{\phi=\phi_f}-V_{\mathrm{tot}}\vert_{\phi=\phi_t}$, $\rho_{\mathrm{rad}}(T)=\pi^2 g_* T^4/30$ and $\phi_{f(t)}$ is the value of the field at the false (true) vacuum.
Another property of a FOPT which is of interest to us is its duration, the inverse of which is defined as,
\begin{align}
\beta/H\approx T_p\left[\dfrac{d}{dT}\left(\dfrac{S_3}{T}\right)\right]_{T=T_p},
\end{align}
which suggests that a larger value of $\beta/H$ denotes a faster FOPT. Another quantity of interest is the reheating temperature, i.e., the temperature where the vacuum energy is converted into radiation and this can be expressed as,
\begin{align}
T_{\mathrm{reh}} = T_p(1+\alpha)^{1/4}.
\end{align}
These are the important parameters of a FOPT which enters and takes pivotal participation in the calculation of the GW amplitude and frequency as well as the mass and abundance of the PBH population which we discuss next.

\subsection{Creation of Gravitational Waves}
\label{sbsc:gwCreatn}
Gravitational wave background of stochastic nature can be created from a FOPT in a few different ways, i.e. (i) collision of bubble walls or relativistic fluid shells, (ii) sound waves in the plasma, and (iii) magnetohydrodynamic turbulence in the background plasma. The first two among these processes has been modeled mathematically with a degree of success, however, the third one is yet to be modelled to a comfortable degree of accuracy. Furthermore, for extremely slow and supercooled FOPTs, there is another source of SGWB, which is the large curvature perturbations (which also creates the PBHs). For our cases, as it is shown in the subsequent sections of this article, there is one supercooled strong FOPT and there is one weaker FOPT driven by a polynomial potential. For the cases of the former FOPT, main contribution for GW will come from collision and curvature perturbations whereas for the later it will come from the sound waves. The brief description for both of these contributions are given as follows.
\subsubsection*{Collision and Curvature Perturbation}
The GW in case of a strongly supercooled and slow FOPT can be expressed as~\cite{Lewicki:2024ghw},
\begin{align}
\Omega_{\mathrm{GW}}^{\mathrm{strong}}h^2 = 1.6\times 10^{-5}\left(\dfrac{g_*}{100}\right)\left(\dfrac{g_{*s}}{100}\right)^{-4/3}\left(\Omega_{\mathrm{coll}}+\Omega_{\mathrm{curv}}\right),
\label{eq:stronggw}
\end{align}
where $g_*$ ($g_{*s}$) is the relativistic (entropy) degrees of freedom at the time of creation of the GW, and $\Omega_{\mathrm{coll}}$ and $\Omega_{\mathrm{curv}}$ are the GW spectrum due to the collision of bubble walls/relativistic fluid shells and the curvature perturbation respectively.

The spectrum for the collision can be expressed as~\cite{Lewicki:2022pdb},
\begin{align}
\Omega_{\mathrm{coll}} = \left(\dfrac{\beta}{H}\right)^{-2}\dfrac{A(a+b)^c S_H(k,k_{\mathrm{max}})}{\left(b(k/k_p)^{-a/c}+a(k/k_p)^{b/c}\right)^c},
\label{eq:omegacoll}
\end{align}
where $k_p\approx 0.7 k_{\mathrm{max}}\beta/H$ is the peak wavenumber, $A = 5.1\times 10^{-2}$ is the amplitude of the spectrum, $a=b=2.4$ and $c=4$~\cite{Lewicki:2022pdb}. Furthermore,
\begin{align}
k_{\mathrm{max}} &= 1.6 \times 10^{-7} \left(\dfrac{g_*}{100}\right)^{1/2}\left(\dfrac{100}{g_{*s}}\right)^{1/3}\dfrac{T_{\mathrm{reh}}}{\mathrm{GeV}}~~~\mathrm{Hz}\\
S_{H}(k,k_{\mathrm{max}}) &= \left(1+\dfrac{\Omega_{\mathrm{CT}}(k_{\mathrm{max}})}{\Omega_{\mathrm{CT}}(k)}\left(\dfrac{k}{k_{\mathrm{max}}}\right)^a\right)^{-1},
\end{align}
where $\Omega_{\mathrm{CT}}$ ($\propto k^3$ in radiation dominated universe) signifies the tail due to the causality~\cite{Lewicki:2022pdb}. 

On the other hand, the spectrum due to the curvature perturbation can be expressed as~\cite{Kohri:2018awv,Espinosa:2018eve,Inomata:2019yww},
\begin{align}
\Omega_{\mathrm{curv}} \approx \dfrac{1}{3}\int_1^{\infty} dt \int_0^1 ds~ \mathcal{I}^2_{t,s}\left(\dfrac{(t^2-1)(1-s^2)}{t^2-s^2}\right)^2\mathcal{P}_{\zeta}\left(k\dfrac{t-s}{s}\right)\mathcal{P}_{\zeta}\left(k\dfrac{t+s}{s}\right),
\label{eq:omegacurv}
\end{align}
where $\mathcal{P}_{\zeta}$ is the curvature perturbation created due to the supercooled slow FOPT and $\mathcal{I}_{t,s}$ signifies the transfer function and can be expressed as,
\begin{align}
\mathcal{I}^2_{t,s} &= \dfrac{288(s^2 + t^2 - 6)}{(t^2 - s^2)^6}\nonumber\\
&\times\left(\dfrac{\pi^2}{4}(s^2 + t^2 - 6)^2\theta(t^2 - 3)+\left(t^2 - s^2 - \dfrac{1}{2}(s^2 + t^2 - 6)\ln\left|\dfrac{t^2-3}{3-s^2}\right|\right)^2\right).
\end{align}
\subsubsection*{Sound Waves}
The SGWB spectra due to sound waves can be expressed as~\cite{Hindmarsh:2013xza, Hindmarsh:2015qta, Hindmarsh:2017gnf},
\begin{align}
\Omega_{\mathrm{sw}}h^2 = 4.13\times 10^{-7}(R_*H_*)\left(1-\dfrac{1}{\sqrt{1+2\tau_{\mathrm{sw}}H_*}}\right)\left(\dfrac{\kappa_{\mathrm{sw}}\alpha}{1+\alpha}\right)^2\left(\dfrac{100}{g_*}\right)^{1/3}S_{\mathrm{sw}}(f),
\end{align}
where,
$S_{\mathrm{sw}}$ contains the information regarding the frequency dependence of the SGWB spectrum and can be expressed as,
\begin{align}
S_{\mathrm{sw}}(f) = \left(\dfrac{f}{f_{\mathrm{sw}}}\right)^3\left(\dfrac{4}{7}+\dfrac{3}{7}\left(\dfrac{f}{f_{\mathrm{sw}}}\right)^2\right)^{-7/2},
\end{align}
the peak frequency of the spectrum is given by,
\begin{align}
f_{\mathrm{sw}} = 2.6\times 10^{-5}(R_*H_*)^{-1}\left(\dfrac{T_{\mathrm{reh}}}{100~\mathrm{GeV}}\right)\left(\dfrac{g_*}{100}\right)^{1/6},
\end{align}
and the other relevant quantities are defined as,
\begin{align}
\tau_{\mathrm{sw}}H_* &= \dfrac{R_*H_*}{U_f},\\
U_f &= \sqrt{\dfrac{3}{4}\dfrac{\alpha}{1+\alpha}\kappa_{\mathrm{sw}}},\\
R_*H_* &= (8\pi)^{1/3}\mathrm{max}(v_w,~c_s)\left(\dfrac{\beta}{H}\right)^{-1}.
\end{align}
Here $c_s = 1/\sqrt{3}$ is the speed of sound in the fluid and $v_w$ is the bubble wall velocity for which we use the approximate form~\cite{Lewicki:2021pgr},
\begin{align}
v_w=
\begin{dcases}
        \sqrt{\dfrac{\Delta V_{\mathrm{tot}}}{\alpha\rho_\mathrm{rad}}}, & \sqrt{\dfrac{\Delta V_{\mathrm{tot}}}{\alpha\rho_\mathrm{rad}}}< v_J, \\
        1, & \sqrt{\dfrac{\Delta V_{\mathrm{tot}}}{\alpha\rho_\mathrm{rad}}} \geq v_J,\\
    \end{dcases}
\end{align}
where, $v_J$ is the Chapman-Jouget velocity and is given by~\cite{Espinosa:2010hh},
\begin{align}
v_J = \dfrac{1}{\sqrt{3}}\dfrac{1+\sqrt{3\alpha^2+2\alpha}}{1+\alpha}.
\end{align}
Finally, for the sound wave efficiency factor $\kappa_{\mathrm{sw}} = 1- \kappa$, we use the expression~\cite{Lewicki:2022pdb},
\begin{align}
\kappa = \mathcal{K}\dfrac{R_{\mathrm{eq}}}{R}\dfrac{\gamma}{\gamma_{\mathrm{eq}}},
\end{align}
where $R$ is the radius of the bubble, $\gamma$ is the Lorentz factor of the bubble wall, $R_{\mathrm{eq}}$ is the radius of the bubble when it does not accelerate anymore, $\gamma_{\mathrm{eq}}$ is the Lorentz factor of the bubble wall when $R=R_{\mathrm{eq}}$ and finally $\mathcal{K}$ is a numerical prefactor which contains information of how much of the energy goes into the surrounding fluid and gets converted into thermal energy. For strongly supercooled FOPTs $\mathcal{K}\rightarrow 1$ and in general cases it can be expressed as,
\begin{align}
\mathcal{K} = \left(1-\dfrac{\alpha_{\infty}}{\alpha}\right)\left(1-\dfrac{1}{\gamma_{\mathrm{eq}}^c}\right),
\end{align}
where $\alpha_{\infty} = \Delta P_{1\rightarrow 1}/\rho_{\mathrm{rad}}$ and $\Delta P_{1\rightarrow 1}$ denotes the pressure difference across the bubble wall due to one to one scattering events. Furthermore, in this study we consider $c=1$. For further insight on these quantities see Refs.~\cite{Lewicki:2022pdb,Ellis:2019oqb}.
\subsection{Creation of Primordial Black Holes}
\label{sbsc:pbhCreatn}
Now we discuss the creation of PBHs from a FOPT. The basic mechanism is that since the nucleation of true vacuum bubble is probabilistic process, there may exist some hubble region in which bubble nucleation has not taken place, i.e., that region is still in false vacuum and dominated by vacuum energy whereas its surrounding regions are in true vacuum and dominated by radiation. As a result, due to difference in the nucleation time, large inhomogeneities can generate which can in turn collapse to form PBH. In this regard, we follow the approach of Ref.~\cite{Lewicki:2024ghw}.

The horizon mass during the collapse of the overdense region can be expressed as,
\begin{align}
M_H = 10^{32}\left(\dfrac{100}{g_*}\right)^{1/2}\left(\dfrac{T_{\mathrm{reh}}}{\mathrm{GeV}}\right)^{-2}~~\mathrm{g}.
\end{align}
The abundance of the PBH created through this mechanism is given by~\cite{Lewicki:2024ghw},
\begin{align}
f_{\mathrm{PBH}} \approx 5.5\times 10^6 \exp\left(-0.064e^{0.806\beta/H}\right)\left(\dfrac{g_*}{g_{*s}}\right)\left(\dfrac{T_{\mathrm{reh}}}{\mathrm{GeV}}\right).
\end{align}
Furthermore, the mass function for these PBHs can be expressed as~\cite{Lewicki:2024ghw},
\begin{align}
\psi(M) \propto (M/M_H)^{1+1/\gamma}\exp\left(-c_1(M/M_H)^{c_2}\right),
\end{align}
where $\gamma$ is the coefficient of the critical scaling law of the PBHs which takes the form $M = \kappa M_k(\delta-\delta_c)^{\gamma}$ where $\kappa$ is the numerical prefactor, $M_k$ is the horizon mass when the perturbation enters the horizon, $\delta$ is the overdensity of the region and $\delta_c$ is the overdensity threshold beyond which a region collapses as the radiation pressure can not balance the overdensity. In this case, fixed values of $\gamma = 0.36$, $\kappa = 4$, and $\delta_c = 0.5$ have been considered. It is also to be noted that the values $c_1$ and $c_2$ have weak dependence on $\beta/H$ and they take approximate values of $1.2$ and $2.7$ respectively for $\beta/H = 8$~\cite{Lewicki:2024ghw}. Therefore, it is trivial to see that in case of a FOPT of this type, the maximum PBHs of a certain population will have a certain mass $M_{\mathrm{PBH,~peak}} \approx 0.928M_H$, which we term as `peak mass'.
%

%
After this brief discussion of possible consequences of a FOPT, we move ahead and show how different parameters of our model affects the different observables, such as GW, PBH, DM etc.

\section{Results}
\label{sec:res}
In this section, we discuss the consequence of our model from different aspects. First we list out the relevant observables and the concerned model parameters. We explicitly show the dependence of some of the observables on a few of these parameters. Finally, we show a few benchmark parameters which are optimal for the ongoing and upcoming experiments.

\subsection{Observables and Relevant Parameters}
The list of observables that may arise from this model and the relevant parameters which effect them are as follows.
\begin{enumerate}
\item \textbf{Gravitational Waves:} Two different set of SGWB may arise from this model, i.e. the high frequency SGWB from the FOPT driven by $\chi$ and the low frequency SGWB from the EWSB if that is a FOPT. The relevant parameters for the high frequency SGWB are $\lambda_4$, the $U(1)_{B-L}$ coupling with $\chi$, $g_{B-L}$; the Yukawa couplings of the right handed neutrinos with $\chi$, $y_{1,2,3}$; and the VEV acquired by $\chi$, $v_{\chi}$. The parameters which affect the low frequency SGWB are $\lambda_{1,2,3,5,6,7,8}$, the Yukawa couplings of the SM Higgs with the SM fermions, the $SU(2)_{L}\times U(1)_{Y}$ gauge coupling of the SM Higgs, and the SM Higgs VEV.
\item \textbf{Primordial Black Holes:} In our model, for the FOPT driven by the two doublets, the $\beta/H$ is much higher than the value required for PBH formation with observationally allowed PBH parameters. Therefore, in this article, the PBH creation is to be accommodated by the FOPT driven by $\chi$. Just like the high frequency SGWB, the PBH properties also depend on $\lambda_4$, $y_{1,2,3}$, $g_{B-L}$, and $v_{\chi}$.
\item \textbf{Particle DM:} The relic density of the particle dark matter, which is the lighter neutral scalar component of the BSM scalars in our model, depends on all the parameters which also affect the phase transition driven by the the SM Higgs field. These parameters are $\lambda_{1,2,3,5,6,7,8}$, and the SM higgs VEV, $v$.
\end{enumerate}

\subsection{Parameter Scan}
We can further divide the parameter scan effectively into two parts, namely, (i) $\chi$-related parameters, and (ii) $\Phi_{1,2}$-related parameters. 

\subsubsection{$\chi$-Related Parameters}
\begin{figure}[t]
\centering
\includegraphics[scale=0.65]{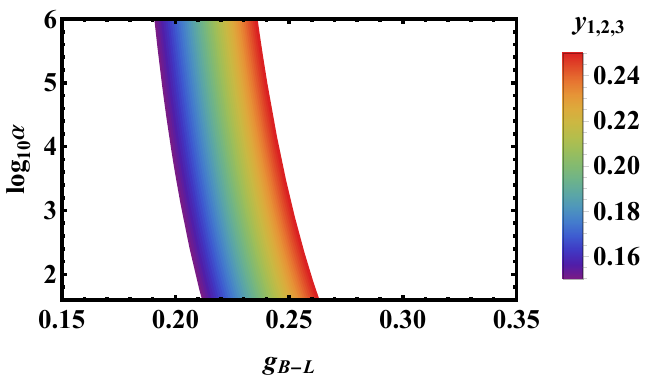}~~~
\includegraphics[scale=0.65]{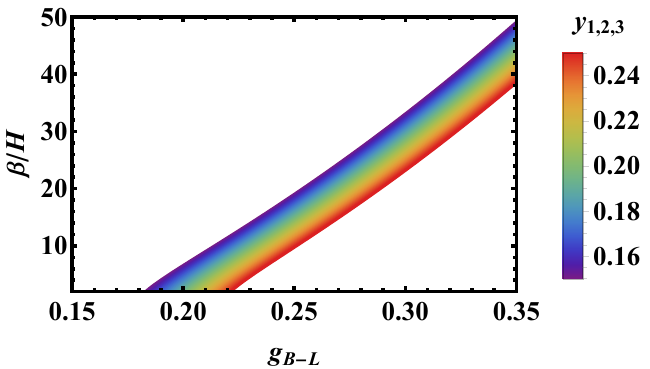}
\caption{Dependence of (left) $\alpha$ and (right) $\beta/H$ values on $g_{B-L}$ for different values of $y_{1,2,3}$. In this case $v_{\chi}=10^8\mathrm{~GeV}$.}
\label{alphabetagy}
\end{figure}

\begin{figure}[t]
\centering
\includegraphics[scale=0.6]{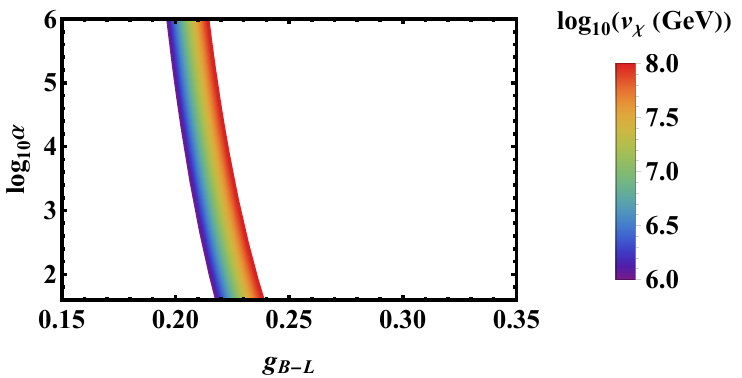}~
\includegraphics[scale=0.6]{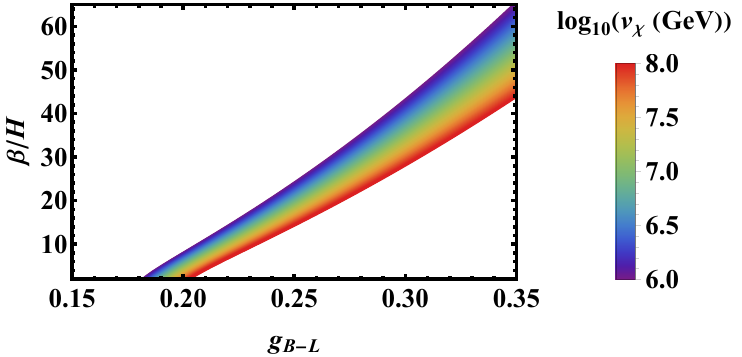}
\caption{Dependence of (left) $\alpha$ and (right) $\beta/H$ values on $g_{B-L}$ for different values of $v_{\chi}$. In this case $y_{1,2,3} = 0.2$.}
\label{alphabetavev}
\end{figure}

In this part, we discuss the parameter ranges relevant for our work to understand the properties of the first FOPT, i.e. the one driven by the scalar field $\chi$. The parameters in question for this case are $g_{B-L}$, $y_{1,2,3}$, and $v_{\chi}$. As mentioned before, these parameters control the strength ($\alpha$), the inverse of the duration ($\beta/H$) and the relevant temperatures ($T_p,~T_n,~T_c$) of the FOPT, and those in turn control the PBH mass and abundance and GW spectrum.
At first in Fig.~\ref{alphabetagy} we show the dependence of $\alpha$ (left panel) and $\beta/H$ (right panel) on the couplings $g_{B-L}$ with varying Yukawa couplings while keeping the value of the VEV of the $\chi$ field fixed at $v_{\chi}=10^8\mathrm{~GeV}$.
It can be seen from Fig.~\ref{alphabetagy} that for a fixed value of $y_{1,2,3}$, $\alpha$ ($\beta/H$) increases (decreases) significantly as $g_{B-L}$ decreases. On the other hand lower value of the Yukawa coupling leads to lower (higher) value of $\alpha$ ($\beta/H$). It is also to be noted that both $\alpha$ and $\beta/H$ depend very weakly on $y_{1,2,3}$ as compared to $g_{B-L}$. This behaviour is also true for the dependence $\alpha$ and $\beta/H$ on the $v_{\chi}$. We illustrate this further in Fig.~\ref{alphabetavev} where we show the dependence of $\alpha$ (left panel) and $\beta/H$ (right panel) on $g_{B-L}$ for different values of VEV for constant Yukawa coupling $y_{1,2,3}=0.2$. 

It can be seen from the figure that for higher value of VEV, $\alpha$ is higher and $\beta/H$ is lower.

\begin{figure}[H]
\centering
\includegraphics[scale=0.6]{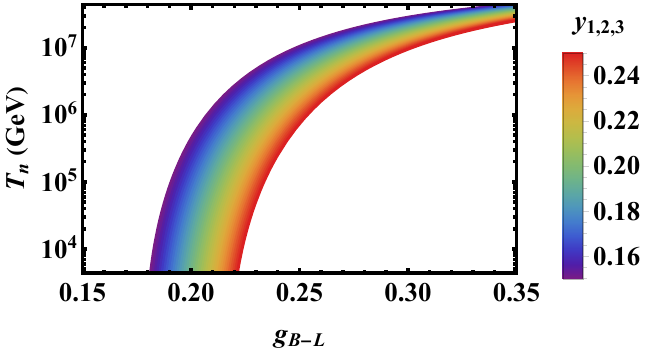}~~~~
\includegraphics[scale=0.6]{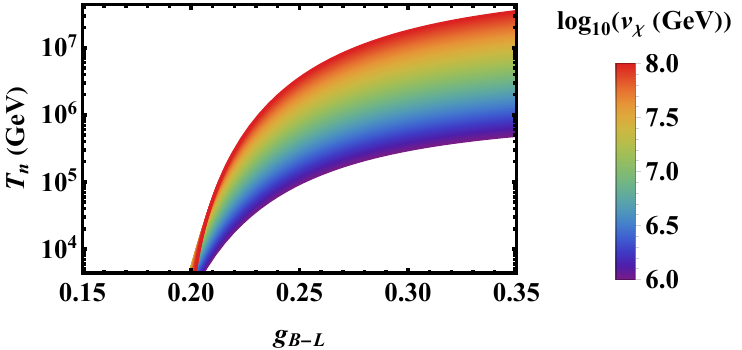}
\caption{Dependence of $T_n$ on $g_{B-L}$ (left) and $y_{1,2,3}$ while keeping $v_{\chi}=10^8\mathrm{~GeV}$ (right) and $v_{\chi}$ keeping $y_{1,2,3}=0.2$.}
\label{Tgy}
\end{figure}
\begin{figure}[H]
\centering
\includegraphics[scale=0.6]{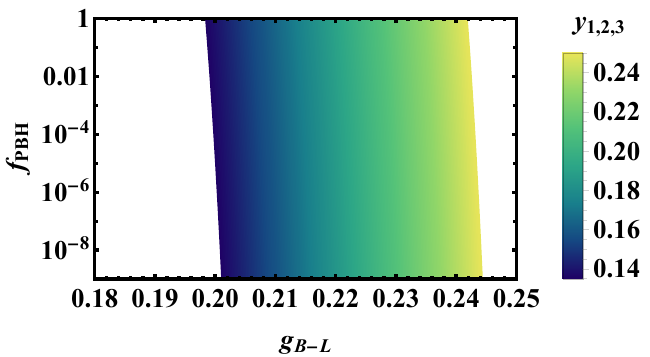}~~~
\includegraphics[scale=0.6]{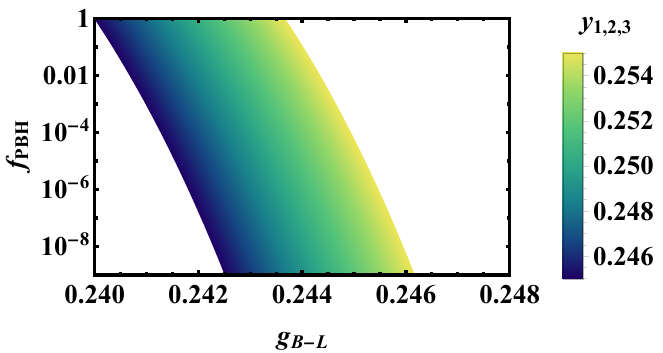}
\caption{(Left) We show the dependence of $f_{\mathrm{PBH}}$ with $g_{B-L}$ at different values of $y_{1,2,3}$. (Right) We show a small region of the left panel plot to further emphasize of the behaviour of $f_{\mathrm{PBH}}$ on varying $g_{B-L}$ and $y_{1,2,3}$. In both the cases $v_{\chi}=10^8\mathrm{~GeV}$.}
\label{fpbhgy}
\end{figure}
\begin{figure}[H]
\centering
\includegraphics[scale=0.6]{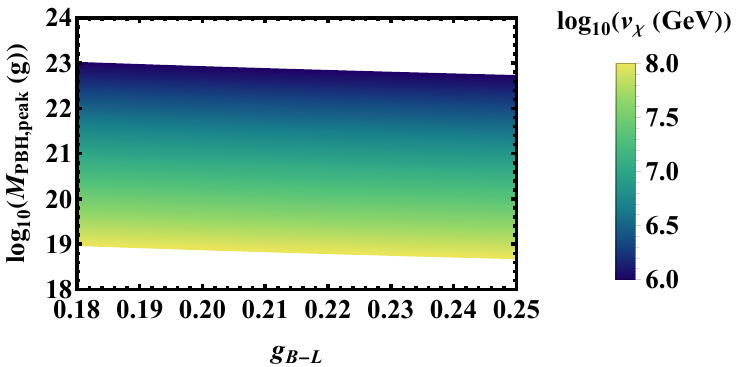}
\caption{We show the dependence of the mass of the PBH generated from the FOPT on the coupling $g_{B-L}$ and the VEV $v_{\chi}$.  In this case $y_{1,2,3} = 0.2$.}
\label{mpbhgy}
\end{figure}
\begin{figure}[H]
\centering
\includegraphics[scale=0.6]{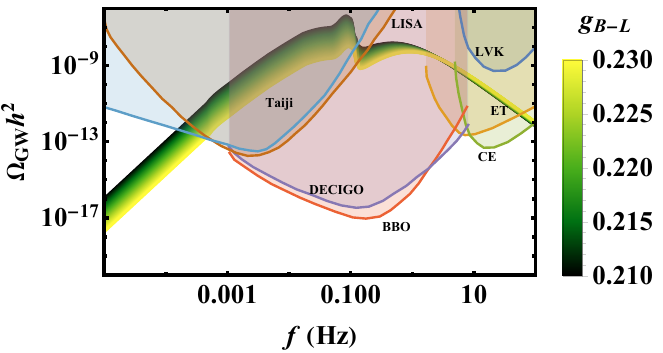}~~~
\includegraphics[scale=0.6]{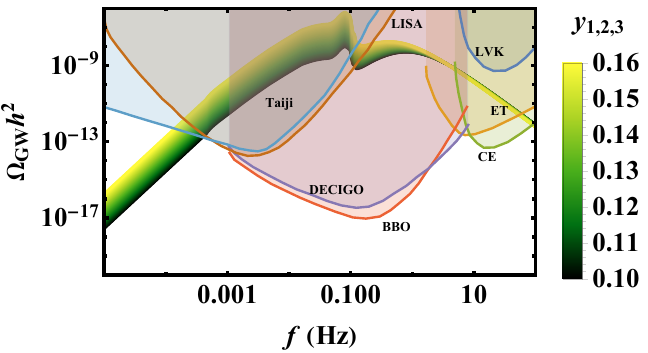}\\
\includegraphics[scale=0.6]{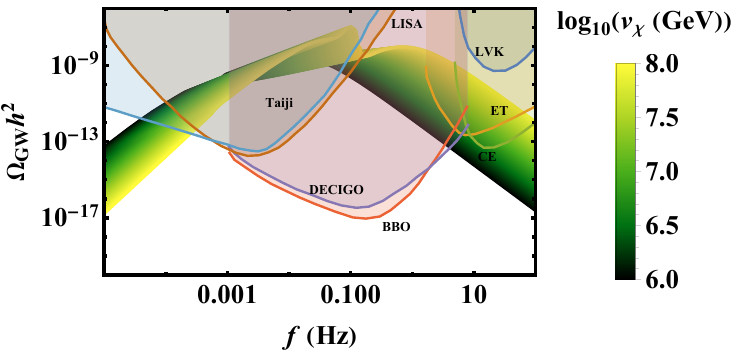}
\caption{(Top-left) We show the SGWB spectra for varying $g_{B-L}$ while keeping $y_{1,2,3}=0.18$ and $v_{\chi}=10^8\mathrm{~GeV}$. (Top-right) We show the SGWB spectra for varying $y_{1,2,3}$ while keeping $g_{B-L} = 0.2$ and $v_{\chi}=10^8\mathrm{~GeV}$. (Bottom) We show the SGWB spectra for varying $v_{\chi}$ while keeping $y_{1,2,3}=0.27$ and $g_{B-L}=0.25$. In both cases we show the sensitivity of the proposed GW detectors such as Taiji~\cite{Ruan:2018tsw}, LISA~\cite{LISA:2017pwj}, LVK~\cite{Somiya:2011np,LIGOScientific:2014pky}, CE~\cite{Reitze:2019iox}, ET~\cite{Punturo:2010zz}, DECIGO~\cite{Kawamura:2011zz}, and BBO~\cite{Phinney:2004bbo}.}
\label{GWchi}
\end{figure}

Apart from $\alpha$ and $\beta/H$, the temperatures related to the FOPT, i.e., $T_p$, $T_n$, and $T_c$ are also important since the peak of the SGWB and the mass of the PBH depends heavily on them. We found that for fixed VEV (Yukawa coupling), all three temperatures have similar behaviour with varying gauge and Yukawa coupling (VEV). Therefore in Fig.~\ref{Tgy} we only show the $T_n$ dependence on $g_{B-L}$ and $y_{1,2,3}$ keeping $v_{\chi}=10^8\mathrm{~GeV}$ (left panel) and $T_n$ dependence on $g_{B-L}$ and $v_{\chi}$ keeping $y_{1,2,3}=0.2$ (right panel).
From the figure, it can be seen that at fixed VEV and Yukawa coupling, $T_n$ increases with increasing $g_{B-L}$. In this case as well, $T_n$ weaker dependence on the Yukawa couplings as compared to $g_{B-L}$. However, for fixed gauge and Yukawa couplings, $T_n$ increases with VEV. 
Now that we have the relevant FOPT parameters as functions of the model parameters, i.e. $g_{B-L}$, $y_{1,2,3}$, and $v_{\chi}$, we can express the properties of the PBH population, i.e., the mass and the abundance as functions of those parameters as well.
In Fig.~\ref{fpbhgy} we show the dependence of the PBH abundance on the couplings with the value of $v_{\chi}=10^8\mathrm{~GeV}$.
From the left panel of the figure, it can be seen that at fixed $g_{B-L}$, extremely small change in the value of $y_{1,2,3}$ can drastically change the value of abundance. The explicit dependence is very hard to comprehend from the left panel of the figure. Therefore, in order to show the strong dependence  of abundance on the couplings, we take a very small parameter range on the right panel of Fig.~\ref{fpbhgy}. It can be seen from the right panel that at fixed $g_{B-L}$ values, $f_{\mathrm{PBH}}$ increases with decreasing $y_{1,2,3}$. Due to this robust dependence, abundance of PBH can be used as a tool to put bounds on such couplings. Next, we focus on the peak mass of the PBH created during the FOPT.
We show the relation of the mass of the PBH and the gauge coupling and the VEV in Fig.~\ref{mpbhgy} keeping the Yuakwa couplings $y_{1,2,3} = 0.2$. It is evident from the figure that the mass of PBH increases with decreasing $g_{B-L}$ and $v_{\chi}$.
Now we focus on the SGWB due to the FOPT driven by $\chi$ for which, as mentioned before, we consider the contributions from the collision of the bubble walls and the curvature perturbation. As mentioned before, the spectrum and the peak frequency of the SGWB depend on the FOPT parameters, i.e., $\alpha$, $\beta/H$, $T_{\mathrm{reh}}$ and those in turn depend on $g_{B-L}$, $y_{1,2,3}$, and $v_{\chi}$. Therefore, in Fig.~\ref{GWchi} we show the dependence on the SGWB spectrum arising from the FOPT on $g_{B-L}$ (top-left panel), $y_{1,2,3}$ (top-right panel), and $v_{\chi}$ (bottom panel) respectively. 
It is evident that at higher values of $g_{B-L}$ the peak frequency increases and the amplitude decreases. Whereas, for increasing $y_{1,2,3}$ the amplitude of the GW spectra increases but the peak frequency remains constant. Finally, for increasing values of $v_{\chi}$  the peak frequency and the amplitude both increases.

\subsubsection{$\Phi_{1,2}$-Related Parameters}
In the context of inert doublet model Ref.~\cite{Benincasa:2022elt} has shown FOPTs driven by both $\Phi_1$ and $\Phi_2$. Here we focus on the possibility of a second FOPT, i.e., the EWSB driven by the doublets. However, the FOPTs which may occur in the physically allowed parameter space are much weaker than the one driven by $\chi$ and hence for simplicity we constrain ourselves in to the FOPTs driven by SM higgs with contributions from the BSM doublet.

In this regard, we express the parameters in our model as functions the quantities that are physically measurable to ensure that the tree-level potential be minimized at the EWSB vacuum in the following manner.
\begin{align}
\lambda_1 &=\dfrac{m_h^2}{2 v^2}, ~~ \lambda_7 =-\dfrac{m_h^2}{2v_{\chi}^2}, ~~    \lambda_8 =\dfrac{2m_{H}^2-\lambda_{356}v^2}{2v_{\chi}^2},   \nonumber\\
\lambda_3 &=\lambda_{356}+ \frac{2}{v^2}\left(m_{H^{\pm}}^2-m_{H}^2\right),~~ \lambda_5=\frac{1}{v^2}\left(m_H^2+m_A^2-2m_{H^{\pm}}^2\right),~~ \lambda_6=\frac{1}{v^2}\left(m_H^2-m_A^2\right),
\end{align}
where $m_{h,H,A,H^{\pm}}$ are the masses of the SM higgs, the extra neutral scalars $H$ and $A$ and the charged scalar $H^{\pm}$ and $\lambda_{356}=\lambda_3+\lambda_5+\lambda_6$. It is worth mentioning here that $H$($A$) will play the role of particle DM if the the mass of $H$ is lower (higher) than $A$. For our parameter choices we find $m_H<m_A$, hence $H$ is the particle DM candidate. From the above relations we can see that since the mass of the SM higgs is well-measured the coupling $\lambda_7$ depends on  $v_{\chi}$. For example, SM higgs mass can be obtained if $v_{\chi}=10^{7}\mathrm{~GeV}$ and $\lambda_7=-7.8125\times 10^{-11}$. Therefore, in order to facilitate the EWSB in our model, the coupling $\lambda_7$ has to be specifically fine-tuned. Although $\lambda_8$ has less constraints on it due to $m_{H}$ and $\lambda_{356}$ being variables, still its order will be determined by $v_{\chi}$ due to its value being much larger than $v$ which has been experimentally determined to be 246 GeV. The values of $\lambda_{3}$, $\lambda_{5}$, and $\lambda_{6}$ are determined by $m_{H,A,H^{\pm}}$ along with the equality that $\lambda_{356}=\lambda_3+\lambda_5+\lambda_6$.
\begin{figure}[t]
	\centering
	\includegraphics[scale=0.33]{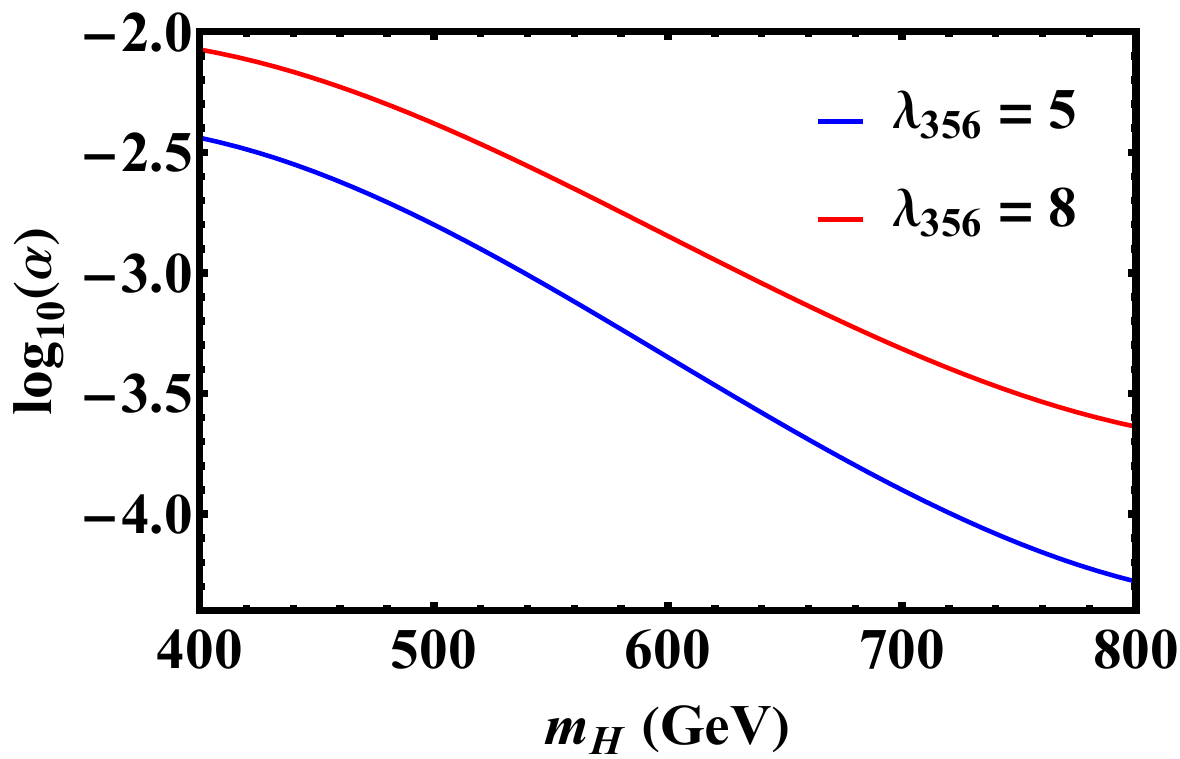}~~~~
	\includegraphics[scale=0.31]{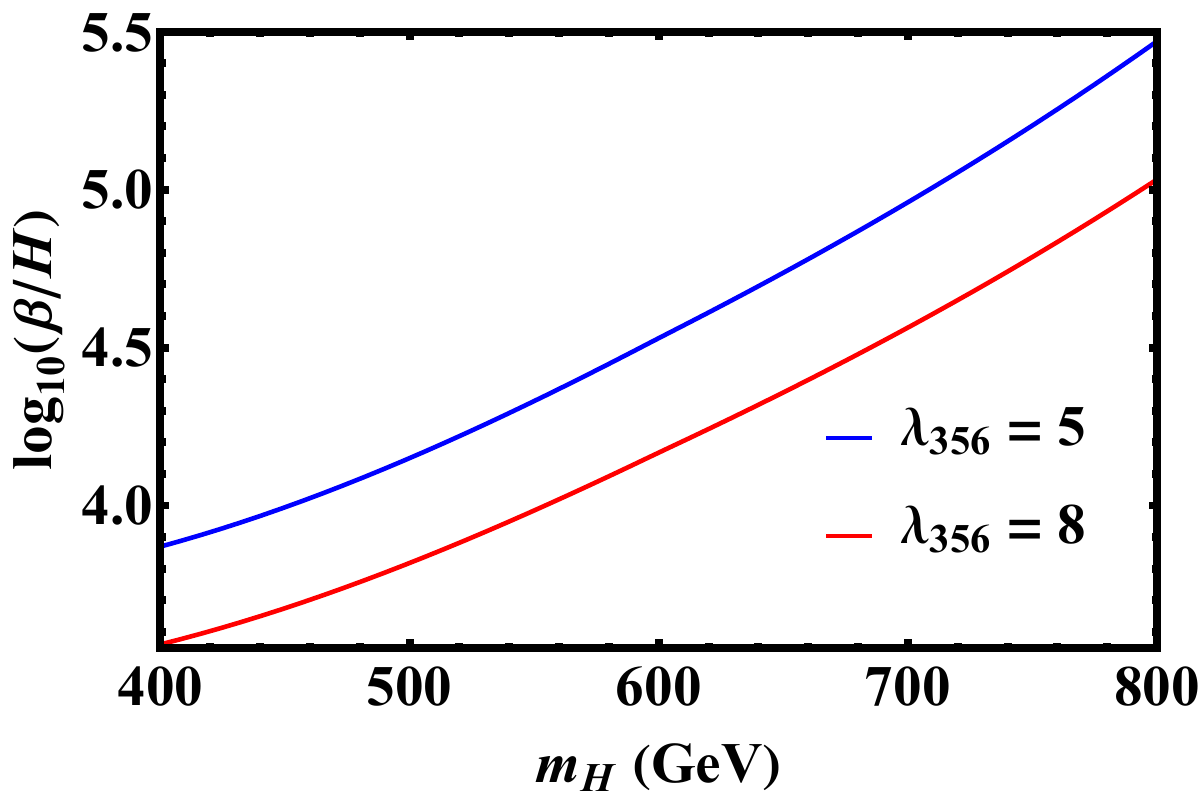}
	\caption{Dependence of (left) $\alpha$ and (right) $\beta/H$ on $m_{H}$ with $\lambda_{356}=5,~8$. In both the cases $m_A=1000\mathrm{~GeV}$, $m_{H^{\pm}}=800\mathrm{~GeV}$ and $\lambda_2=2$.}
	\label{alphabetaewsb}
\end{figure}
It has been shown that the mass of the DM candidate has to be larger than $\mathcal{O}(60)\mathrm{~GeV}$ for FOPTs to happen in the single step configuration~\cite{Benincasa:2022elt}. 
Moreover, to satisfy the relic density constraint one has to resort to only the resonance region i.e., $\sim 62.5$ GeV. Since in our model we can have PBH as the primary DM candidate we relax the constricted mass range for the scalar DM 
and take $m_{H}\in [200,1000]\mathrm{~GeV}$. Furthermore, it is also shown that $\lambda_2$ also does not have a strong dependence on FOPT. However, $\lambda_{356}$ has a strong dependence on the possibility of a FOPT. Therefore, remaining within the parameter space verified in Ref.~\cite{Benincasa:2022elt}, we show the dependence of $\lambda_{356}$ and $m_{H}$ on the FOPT parameters for completeness.
It can be seen from Fig.~\ref{alphabetaewsb}, for the entire parameter space, the $\alpha$ (left panel) and $\beta/H$ (right panel) are very small and very large respectively. Furthermore, it can also be seen that both $\alpha$ and $\beta/H$ has much stronger dependence on $m_H$ as compared to $\lambda_{356}$. It is worth mentioning here that for  this FOPT, we find $T_n\sim T_p$. For the entire parameter space that we consider, $T_n\in[156,199]\mathrm{~GeV}$.
\begin{figure}[t]
\centering
\includegraphics[scale=0.65]{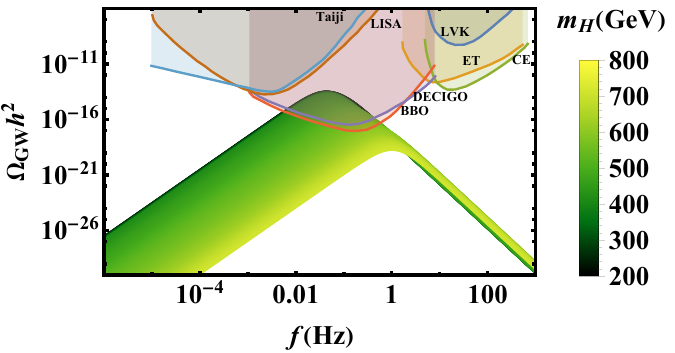}~~~
\includegraphics[scale=0.65]{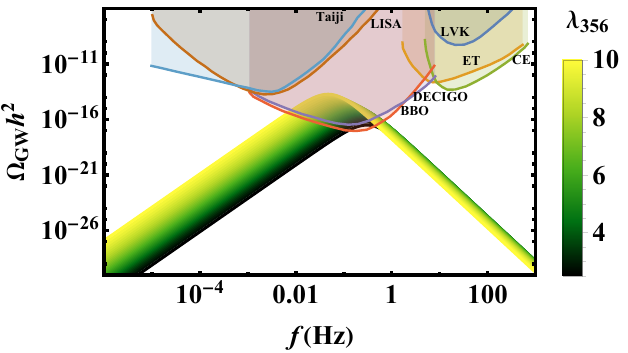}
\caption{Dependence of the GW spectrum from the EWSB on (left) $m_H$ and (right) $\lambda_{356}$. In the left panel $\lambda_{356}=10$ and in the right panel $m_{H}=450\mathrm{~GeV}$. In both the cases $m_A=1000\mathrm{~GeV}$, $m_{H^{\pm}}=800\mathrm{~GeV}$ and $\lambda_2=2$.}
\label{GWewsb}
\end{figure}
In Fig.~\ref{GWewsb} we show the dependence of the SGWB spectrum originating from this FOPT on $m_H$ (left panel) and $\lambda_{356}$ (right panel). It can be seen from the figure that as $\lambda_{356}$ increases and $m_{H}$ decreases the amplitude of the SGWB spectra increases. In our parameter space for $\lambda_{356}=10$ and $m_{H}=200\mathrm{~GeV}$ we obtain the strongest SGWB signal which can be detected in DECIGO and BBO. However, it should be noted that the $\chi$ driven FOPT creates much larger SGWB, which will overshadow the SGWB due to the $\Phi_{1,2}$ driven FOPT.
\subsection{Benchmark Parameters and Consequences}
\begin{figure}[t]
\centering
\includegraphics[scale=0.6]{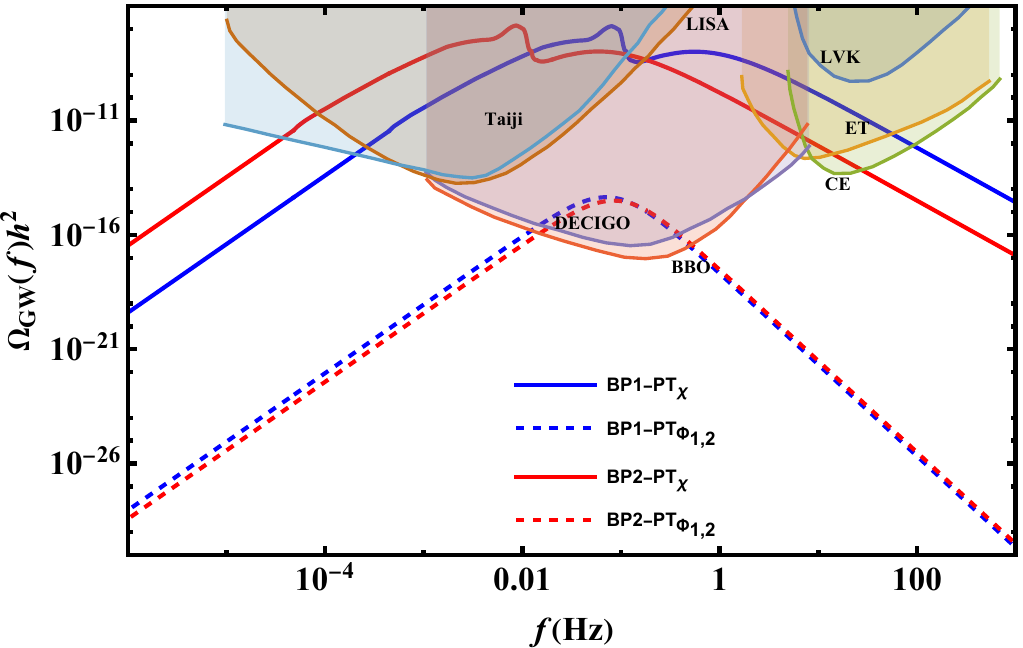}
\caption{The GW spectrum resulting from the two benchmark cases shown in Tab. 1. Relevant sensitivity curves have also been shown.}
\label{BPGW}
\end{figure}

In this part, we take two benchmark cases to explicitly show the consequences of our model. In Tab.~\ref{bpin} we show the two sets benchmark values for the parameters in our model. It should be noted here that we have replaced $\lambda_{3,5,6,8}$ with $m_{H,A,H^{\pm}}$, and $\lambda_{356}$.
\begin{table}[H]
\begin{tabular}{|c|c|c|c|c|c|c|c|c|c|}
\hline
BP&$g_{B-L}$ & $y_{1,2,3}$ & $v_{\chi}(\mathrm{GeV})$ & $\lambda_4$ & $\lambda_2$ & $m_H(\mathrm{GeV})$ & $m_A(\mathrm{GeV})$ & $m_{H^{\pm}}(\mathrm{GeV})$ & $\lambda_{356}$ \\ \hline
 1  & 0.1988   &  0.1356    &   $10^8$       &    $10^{-10}$         &       2      &    500   &   1000    &      800         &       10          \\ \hline
 2   & 0.2154     &     0.2149        &    $10^7$        &      $10^{-10}$       &     3        &   300    &   600    &     1000          &        5         \\ \hline
\end{tabular}
\caption{The benchmark parameters for the two cases.}
\label{bpin}
\end{table}
In Tab.~\ref{bpout} we show the consequences of the benchmark cases considered in Tab.~\ref{bpin}. It can be seen that in both the cases, PBHs effectively contributed to the entire DM fraction of the universe. Furthermore, each of these benchmark cases allows for two FOPTs at different temperatures, hence contributes to two different SGWB spectra which we show in Fig.~\ref{BPGW}. It is to be noted that for the SGWB due to the $\chi$ driven FOPT, the left peak is a result of the curvature perturbation whereas the right peak is a result of the collision of bubble walls. It can be seen that for each case, the FOPT driven by $\chi$ creates SGWB spectra which is in the sensitivity range of LISA, Taiji, DECIGO, BBO, CE, and ET. However, in both the cases the SGWB created during the FOPT driven by the doublets are only within the sensitivity ranges of the detectors BBO and DECIGO. It is worth mentioning that in both these cases the relic density of the particle DM is $\mathcal{O}(10^{-4})$ and PBH plays the role of primary DM in the universe.
\begin{table}[H]
\centering
\begin{tabular}{|c|c|c|c|c|}
\hline
BP&$f_{\mathrm{PBH}}$ & $M_{\mathrm{PBH}}(\mathrm{g})$ & $\mathrm{PT}_{\chi}(\alpha,~\beta/H,~T_{\mathrm{reh}}~(\mathrm{GeV}))$ & $\mathrm{PT}_{\Phi_{1,2}}(\alpha,~\beta/H,~T_n~(\mathrm{GeV}))$\\ \hline
    1  & 0.9997            &        $7.98\times 10^{18}$            &      $(289,~7.65,~3.66\times 10^6)$                &      $(0.013,~3905,~161)$                                     \\ \hline
     2  & 0.9999           &         $6.885\times 10^{20}$           &           $(80908,~7.56,~3.94\times 10^5)$                          &             $(0.0088,~4850,~176.2)$ \\ \hline
\end{tabular}
\caption{The consequences of the benchmark cases.}
\label{bpout}
\end{table}

\section{Conclusion}
\label{sec:concl}
In this article, we have studied a $U(1)_{B-L}$ extension of inert doublet model. In addition to the SM Higgs ($\Phi_1$), we also have another additional (inert) doublet ($\Phi_2$) and a singlet scalar ($\chi$). The singlet scalar has Yukawa coupling with three right handed neutrinos. We further have a gauge boson which realises its mass from its coupling with the singlet scalar as the singlet scalar acquires a VEV $v_{\chi}$. It is to be noted that we consider the Higgs VEV $v \ll v_{\chi}$. 
In this model, for certain part of the parameter space, we can have two first order phase transitions, one high temperature FOPT driven by the scalar $\chi$ and the low temperature FOPT driven by the $\Phi_{1,2}$. Along with the SM Higgs boson, the two doublets also give us one charged scalar $H^{\pm}$ and two neutral scalars $H,~A$ among which the lighter one can play the role of particle DM. However, for a very narrow part of the parameter space the particle DM can have an appreciable relic density. Since in our model, the high temperature phase transition can also create PBH in the mass range which can play the role of the DM entirely, we do not consider the particle DM. Apart from the PBH, both the FOPTs can also generate SGWB spectra some of which can be detected in the future detectors.
In order to quantify the properties of the FOPTs we scan different parameter spaces. We find that for the high temperature FOPT, the strength ($\alpha$) decreases sharply with the $U(1)_{B-L}$ gauge coupling $g_{B-L}$. However, $\alpha$ has weaker dependence on the Yukawa couplings $y_{1,2,3}$ and the $v_{\chi}$. The inverse duration of the FOPT, $\beta/H$ increases sharply with $g_{B-L}$, and similar to the previous case, it has weaker dependence on $y_{1,2,3}$ and $v_{\chi}$. We further find that the critical, nucleation, and percolation temperatures increase with $g_{B-L}$ and $v_{\chi}$ and they have a weaker dependence on $y_{1,2,3}$. For the case of PBH, we found that the abundance is extremely sensitive to the gauge and Yukawa couplings, i.e., for very small increase of the couplings, the abundance decreases drastically. Furthermore, in the coupling range where formation of PBH is allowed, the PBH mass decreases with increasing coupling. The parameter space we consider, generates SGWB spectra is at the detectable range of the future experiments LISA, Taiji, DECIGO, BBO, ET, CE, etc. For the second phase transition we just vary the mass of $H$ and $\lambda_{356}$ to see their effect on the FOPT parameters. We find that $\alpha$ ($\beta/H$) decreases (increases) with increasing mass of $H$, and both of them has very weak dependence on $\lambda_{356}$. We also see that only for a fraction of the parameter range there can be detectable SGWB spectra. We see that the amplitude of the SGWB increases with decreasing $m_{H}$ and increasing $\lambda_{356}$. 
Next, we consider two sets of benchmark values and see their consequences in the PBH and SGWB aspect. We find that though in both the cases there is negligible relic density of the particle DM, the PBH plays the role of entire DM. We also find that although in both cases SGWB from both the FOPTs can be detected by future GW detectors, in both the cases the SGWB from the $\chi$ driven FOPT completely overshadows the other one.
It is worth mentioning here that there are a few conditions and approximations that we have used throughout the article. The first one is that the strong $\chi$ driven phase transition only happens when the self-interaction of $\chi$ is negligible. 
We also have assumed equal Yukawa couplings between the three right handed neutrinos and the singlet scalar. Though for our purpose the unequal couplings would not have changed the results, it can still have other phenomenological implications which we do not focus on. Finally, it is to be noted that different prescriptions available in different literature regarding the creation of PBH during a FOPT and its abundance suggests different values of upper bound of $\beta/H$ for which PBHs can be created. However, we follow one of them since our motivation is to show that various models which have tight constraints on them from the DM aspect, can be revived by using a $U(1)$ extension which can give us FOPTs strong and slow enough to create PBHs with appreciable abundance.
In conclusion, we claim that these kind of extensions can be applied to other models as well which have rich phenomenological implications but are heavily constrained from the DM relic density aspect. 

\acknowledgments{IKB acknowledges the support by the MHRD, Government of India, under the Prime Minister's Research Fellows (PMRF) Scheme, 2022. The work of SK is partially supported by Science, Technology \& Innovation Funding Authority (STDF) under grant number 48173.}


\bibliographystyle{JHEP}
\bibliography{u12hdm.bib}

\end{document}